\documentclass{PoS}

\title{On neutrino production of a charmed meson}

\ShortTitle{Neutrino production of a charmed meson}

\author{\speaker{J. Wagner}\\
        National Centre for Nuclear Research (NCBJ), 00-681 Warsaw, Poland\\
        E-mail: \email{jakub.wagner@ncbj.gov.pl}}

\author{B.~Pire\\
 Centre de Physique Th\'eorique, \'Ecole Polytechnique,
CNRS, Universit\'e Paris-Saclay, 91128 Palaiseau,     France }

\author{L.~Szymanowski\\
 National Centre for Nuclear Research (NCBJ), 00-681 Warsaw, Poland }

\abstract{We calculate in the framework of the collinear QCD approach the  amplitude for exclusive  neutrino-production of a pseudoscalar charmed $D$ meson.  This process allows to access gluon and both  chiral-odd and chiral-even quark generalized parton distributions (GPDs), which contribute in specific ways to the amplitude for different polarization states of the $W$ boson. The energy dependence of the cross section allows to separate different contributions and the measurement of the azimuthal dependence helps to single out the transversity chiral-odd GPDs contributions. The flavor dependence, and in particular the difference between $D^+$ and $D^0$ production rates, allows to test the importance of gluonic contributions. The behaviour of the proton and neutron target cross sections enables to separate the $u$ and $d$ quark contributions. Planned medium and high energy neutrino facilities will thus allow  some important progress in the realm of hadronic physics. }

\FullConference{XXV International Workshop on Deep-Inelastic Scattering and Related Subjects\\
		3-7 April 2017\\
		University of Birmingham, UK}

\begin{document}

\begin{figure}
\includegraphics[width=0.45\textwidth]{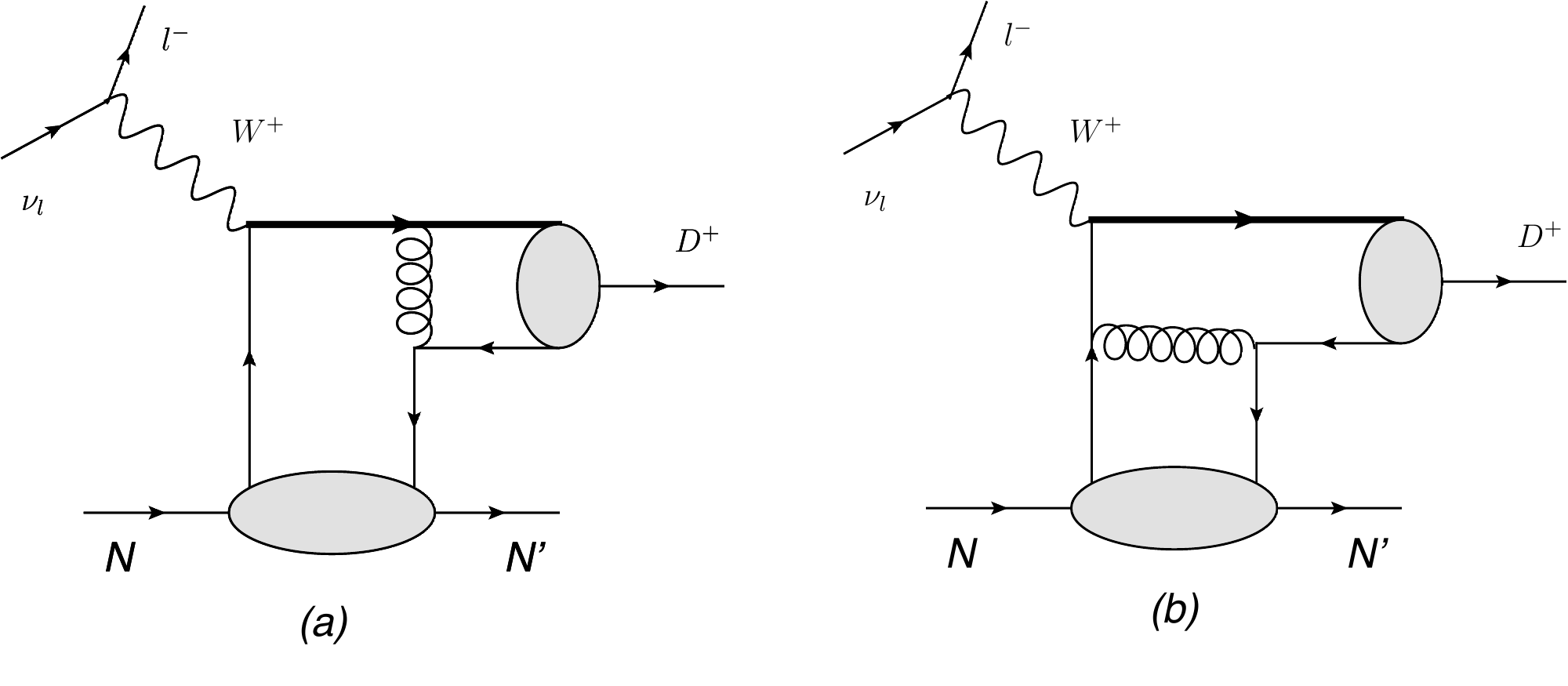}~~~~~~~~~\includegraphics[width=0.45\textwidth]{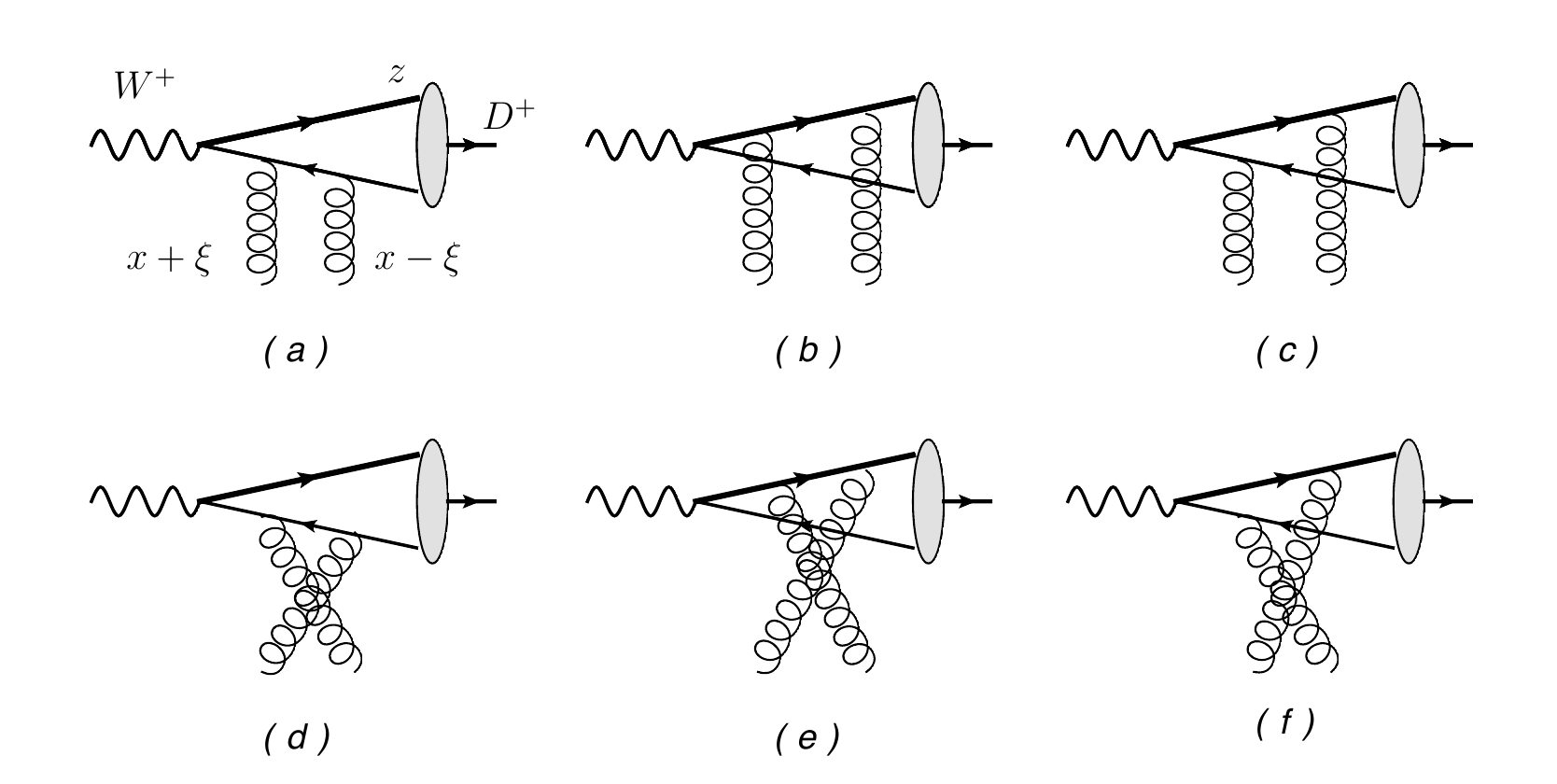}
\caption{Feynman diagrams for the factorized  amplitude for the $ \nu_l N \to l^-  D^+ N'$ or the $ \nu_l N \to l^-  D^0 N'$ process involving the quark GPDs (left panel) or the gluon GPDs (right panel); the thick line represents the heavy quark. }
   \label{Fig1}
\end{figure}

Neutrino production is an interesting way to access (generalized) parton distributions  \cite{weakGPD,Kopeliovich:2012dr} since they allow  to scrutinize the flavor content of the nucleon. In Ref.\cite{Pire:2017lfj} we calculate the exclusive production of a pseudoscalar charmed $D-$meson in the scattering of a neutrino on a proton  or a neutron  target,
in the kinematical domain where collinear factorization  leads to a description of the scattering amplitude 
in terms of nucleon GPDs and the $D-$meson distribution amplitude (DA).  The hard subprocesses at work are:
\begin{eqnarray}
W^+ d \to D^+ d~~~&,&~~~W^+ d \to D^0 u\,,
\end{eqnarray}
described by the  handbag Feynman diagrams of the left panel of Fig. 1 convoluted with chiral-even or chiral-odd quark GPDs, and the hard subprocesses shown on the right panel of Fig. 1:
\begin{eqnarray}
W^+ g \to D^+ g\,,
\end{eqnarray}
convoluted with gluon GPDs.

The new feature of charmed meson neutrino production is the non-zero contribution to the  transverse amplitude of chiral-odd GPDs  through the quark mass effects in the coefficient function  \cite{PS} and the meson mass effect in the DA, which is defined as
\begin{eqnarray}
\langle D^+(P_D) | \bar c_\beta(y) d_\gamma(-y) |0 \rangle & =&
   i \frac{f_D}{4} \int_0^1 dz e^{i(z-\bar z)P_D.y} [(\hat P_D-M_D)\gamma^5]_{\gamma \beta}   \phi_D(z)\,,
   \label{DA}
         \end{eqnarray}
with $z=\frac{k^+}{P_D^+}$ and  where $\int_0^1 dz ~ \phi_D(z) = 1$,  $f_D= 0.223$ GeV. As usual, we denote $\bar z=1-z$ and  $\hat p = p_\mu \gamma^\mu$ for any vector $p$; 
         $ \phi_D(z)$ is  peaked around $z_0 = \frac {m_c}{M_D}$.
The transverse amplitude reads ($\tau = 1-i2$):
\begin{eqnarray}
T_{T} & = &\frac{- i 2C_q \xi (2M_D-m_c)}{\sqrt 2 (Q^2+M_D^2)}  \bar{N}(p_{2}) \left[  {\mathcal{H}}_{T} i\sigma^{n\tau} +\tilde {\mathcal{H}}_{T}\frac{\Delta^{\tau}}{m_N^2}    + {\mathcal E}_{T} \frac{\hat n \Delta ^{\tau}+2\xi  \gamma ^{\tau}}{2m_N} - \tilde {\mathcal E}_{T}\frac{\gamma ^{\tau}}{m_N}\right] N(p_{1}), 
\end{eqnarray}
with $C_q= \frac{2\pi}{3}C_F \alpha_s V_{dc}$, in terms of  transverse form factors that we define as  :
\begin{eqnarray}
{\cal F }_T=f_{D}\int \frac{\phi_D(z)dz}{\bar z}\hspace{-.1cm}\int \frac{F^d_T(x,\xi,t) dx }{(x-\xi+\beta \xi+i\epsilon) (x-\xi +\alpha \bar z+i\epsilon)},
\label{TFF}
 \end{eqnarray} 
where $F^d_T$ is any d-quark transversity GPD, $\alpha = \frac {2 \xi M_D^2}{Q^2+M_D^2}$, $\beta =  \frac {2 (M_D^2-m_c^2)}{Q^2+M_D^2}$.

\begin{figure}
\includegraphics[width=0.45\textwidth]{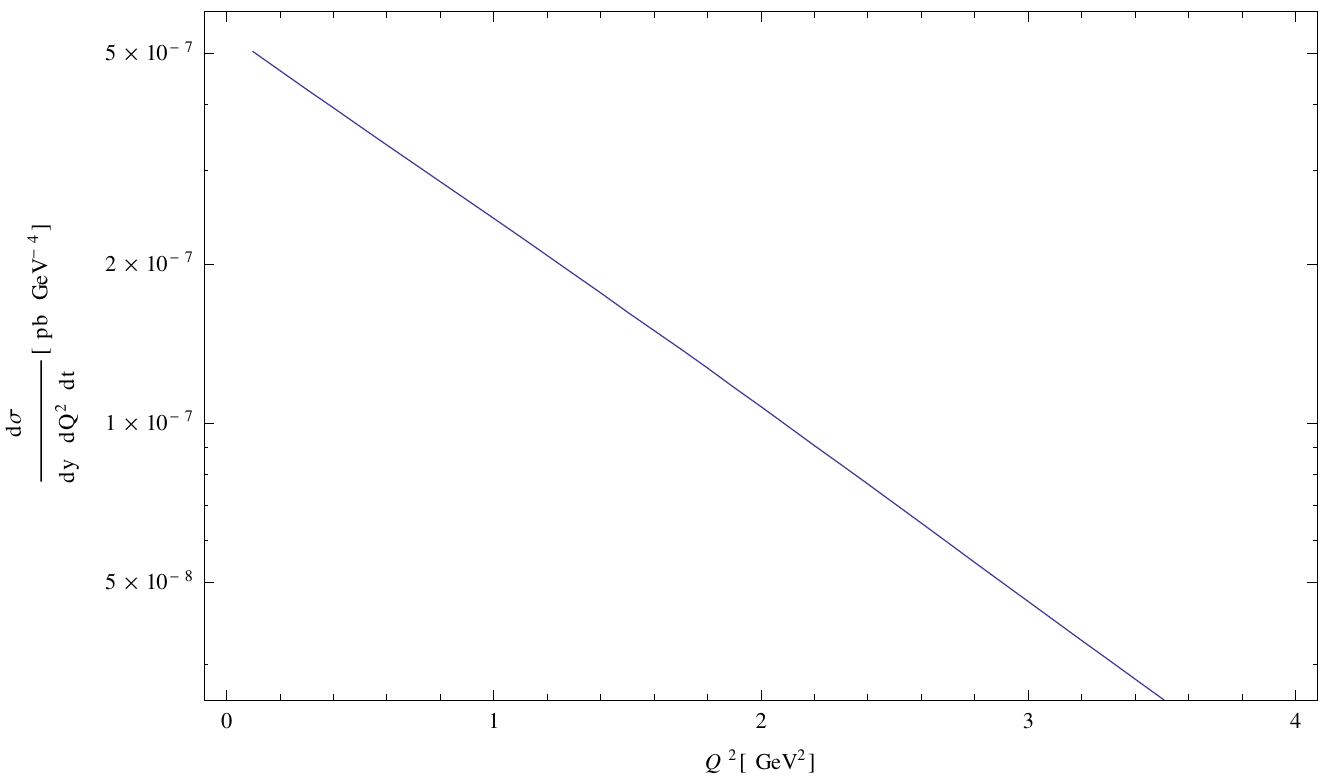} ~~~~\includegraphics[width=0.45\textwidth]{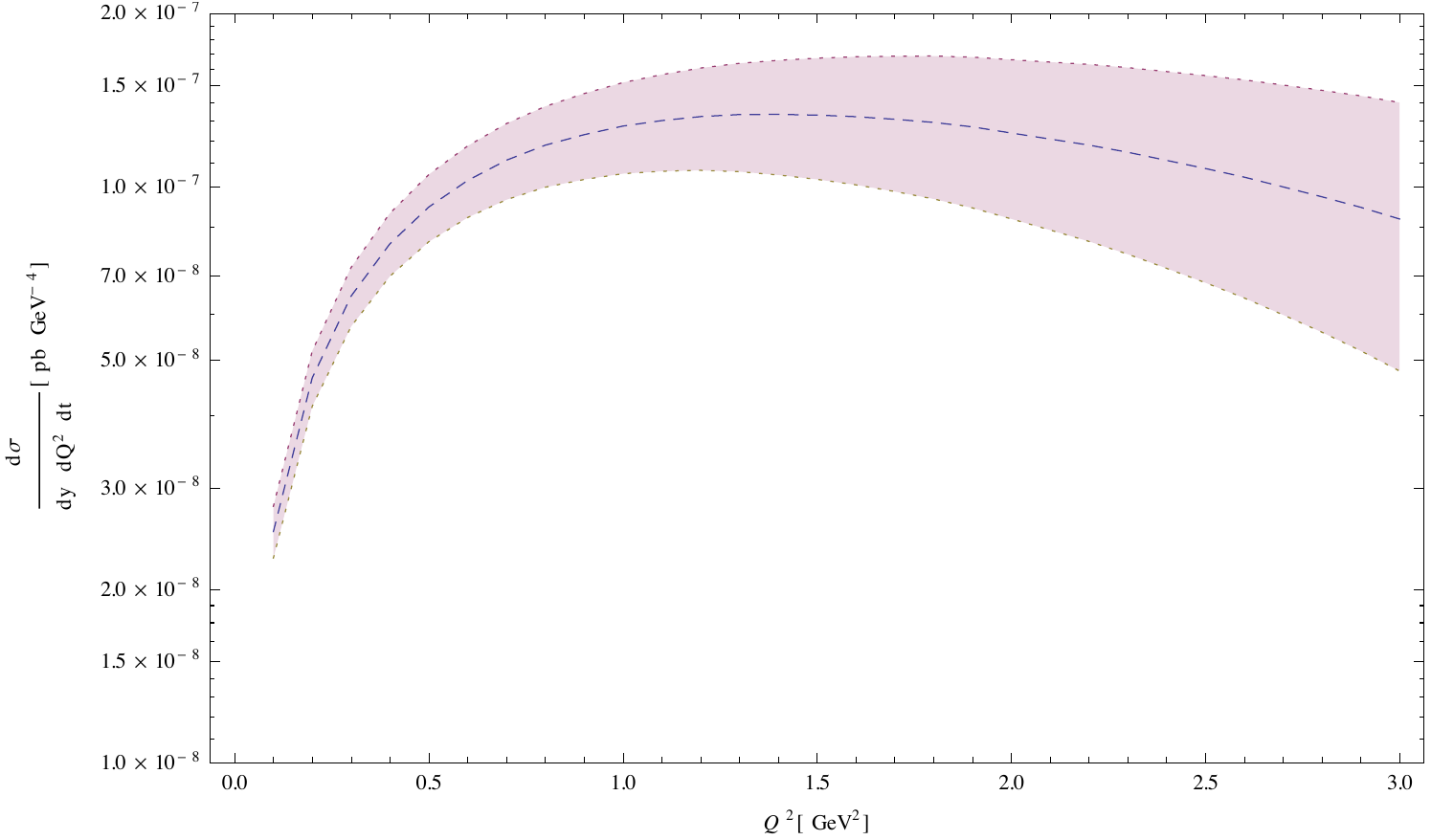}
\caption{The $Q^2$ dependence of the longitudinal (left panel)  and transverse (right panel) contributions to the cross section $\frac{d\sigma(\nu n \to l^- p D^0)}{dy\, dQ^2\, dt}$ (in pb GeV$^{-4}$) for $y=0.7, \Delta_T = 0$  and $s=20$ GeV$^2$. The  band corresponds to  three plausible models  \cite{Pire:2017lfj} for the transversity GPDs.}
   \label{sigma{00}_dy_D0}
\end{figure}

 The differential cross section for neutrino production of a pseudoscalar charmed boson is written as:
 \begin{eqnarray}
\label{cs}
&&\frac{d^4\sigma(\nu N\to l^- N'D)}{dy\, dQ^2\, dt\,  d\varphi} = \\
&&\bar\Gamma
\Bigl\{ ~\frac{1+ \sqrt{1-\varepsilon^2}}{2} \sigma_{- -}+\varepsilon\sigma_{00}+  \sqrt{\varepsilon}(\sqrt{1+\varepsilon}+\sqrt{1-\varepsilon} )(\cos\varphi\
{\rm Re}\sigma_{- 0} + \sin\varphi\
 {\rm Im}\sigma_{- 0} )\ \Bigr\}, \nonumber
\end{eqnarray}
with $y= \frac{p \cdot q}{p\cdot k}$ , $Q^2 = x_B y (s-m^2)$, $\varepsilon \approx \frac{1-y}{1-y+y^2/2}$ and
$
\bar \Gamma = \frac{G_F^2}{(2 \pi)^4} \frac{1}{32y} \frac{1}{\sqrt{ 1+4x_B^2m_N^2/Q^2}}\frac{1}{(s-m_N^2)^2} \frac{Q^2}{1-\epsilon}\,, \nonumber
$
where the ``cross-sections'' $\sigma_{lm}=\epsilon^{* \mu}_l W_{\mu \nu} \epsilon^\nu_m$ are product of  amplitudes for the process $ W(\epsilon_l) N\to D N' $, averaged  (summed) over the initial (final) hadron polarizations. Integrating over $\varphi$ yields the differential cross section :
 \begin{eqnarray}
\label{csphi}
\frac{d\sigma(\nu N\to l^- N'D)}{dy\, dQ^2\, dt}
 = 2 \pi \bar\Gamma
\Bigl\{ ~\frac{1+ \sqrt{1-\varepsilon^2}}{2} \sigma_{- -}&+&\varepsilon\sigma_{00} \Bigr\}.
\end{eqnarray} 

\begin{figure}
\includegraphics[width=0.8\textwidth]{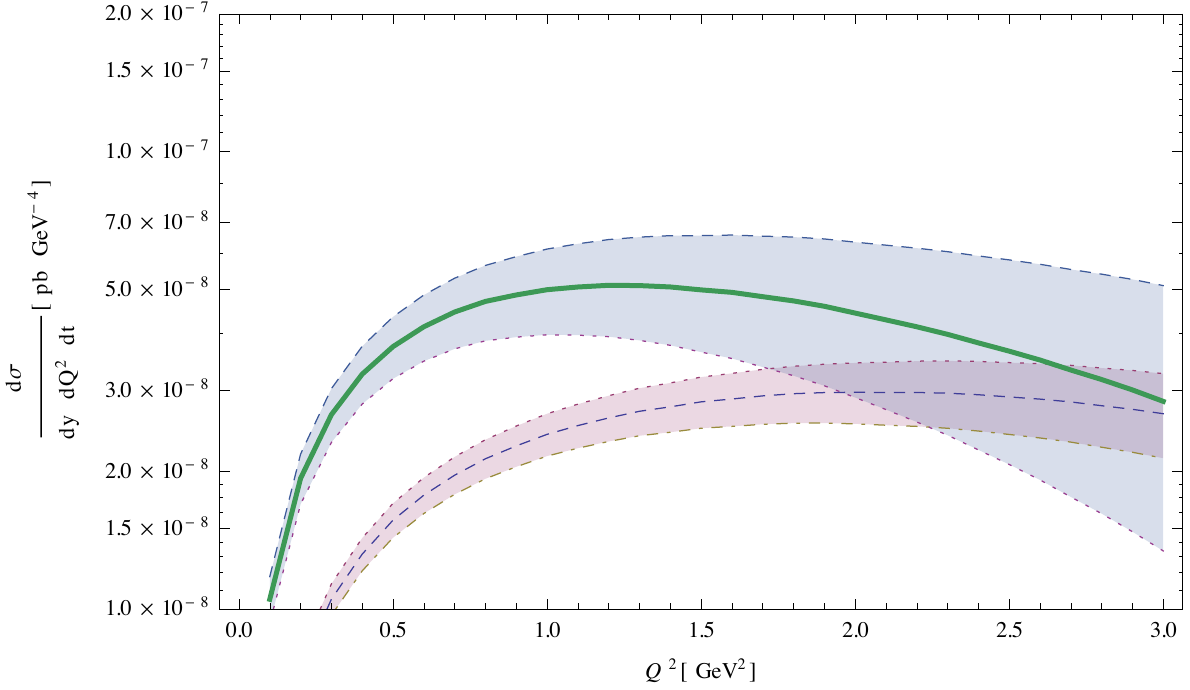}
\caption{The $Q^2$ dependence of the transverse contribution to the cross section $\frac{d\sigma(\nu N \to l^- N D^+)}{dy\, dQ^2\, dt}$ (in pb GeV$^{-4}$) for $y=0.7, \Delta_T = 0$  and $s=20$ GeV$^2$ for a proton (dashed curve and lower band) and neutron (solid curve and upper band) target. The  bands correspond to  three plausible models  \cite{Pire:2017lfj} for the transversity GPDs.}
   \label{sigma{--}_dy_D+_on_proton_and_neutron}
\end{figure}

\begin{figure}
\includegraphics[width=0.95\textwidth]{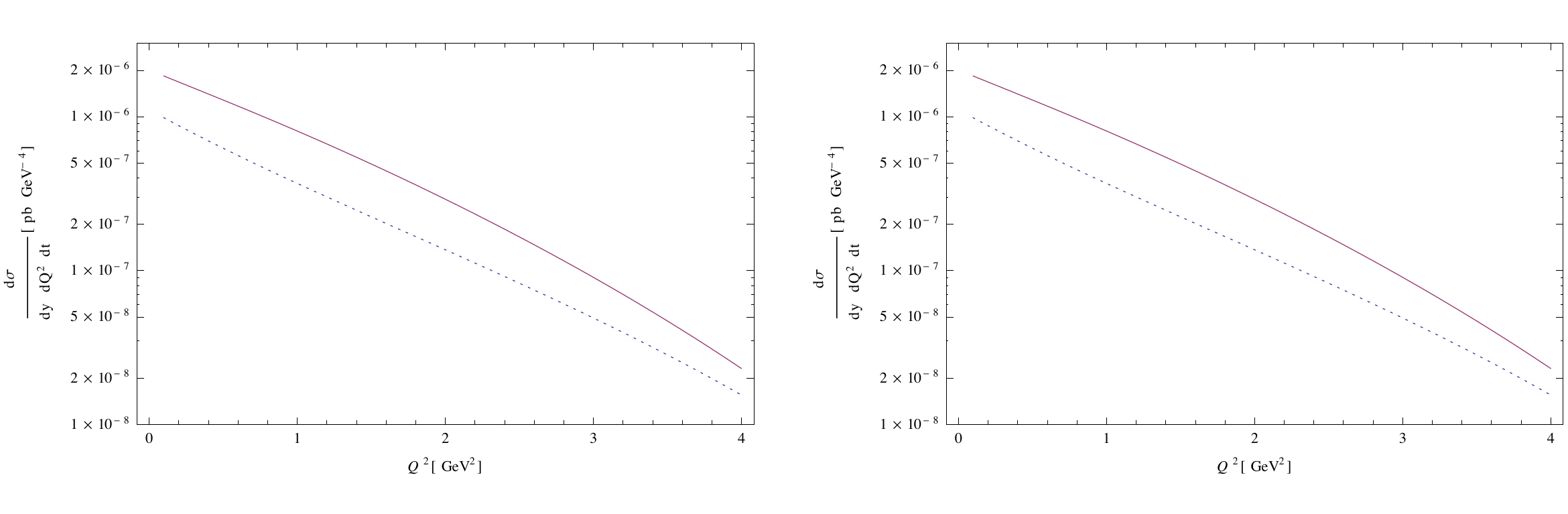}
\caption{The $Q^2$ dependence of the quark (dashed curve) contribution compared  to the total (quark and gluon, solid curve) longitudinal cross section $\frac{d\sigma(\nu N \to l^- N D^+)}{dy\, dQ^2\, dt}$ (in pb GeV$^{-4}$) for $D^+$ production  on a proton (left panel) and neutron (right panel) target for $y=0.7, \Delta_T = 0$  and $s=20$ GeV$^2$.}
   \label{sigma{00}_dy_D+_on_proton_neutron_quarks_gluons}
\end{figure}

Gluon GPDs do not contribute to $D^0$ production since the flavor of the baryon is changed in this reaction. We show on Fig.  \ref{sigma{00}_dy_D0} the longitudinal and  transverse contributions to the cross section.
The transverse contribution is noteworthy of the same order of magnitude as the longitudinal one and even dominates for $y$ large enough. Accessing the chiral-odd transversity GPDs indeed seems feasible in this reaction.

\begin{figure}
\includegraphics[width=0.45\textwidth]{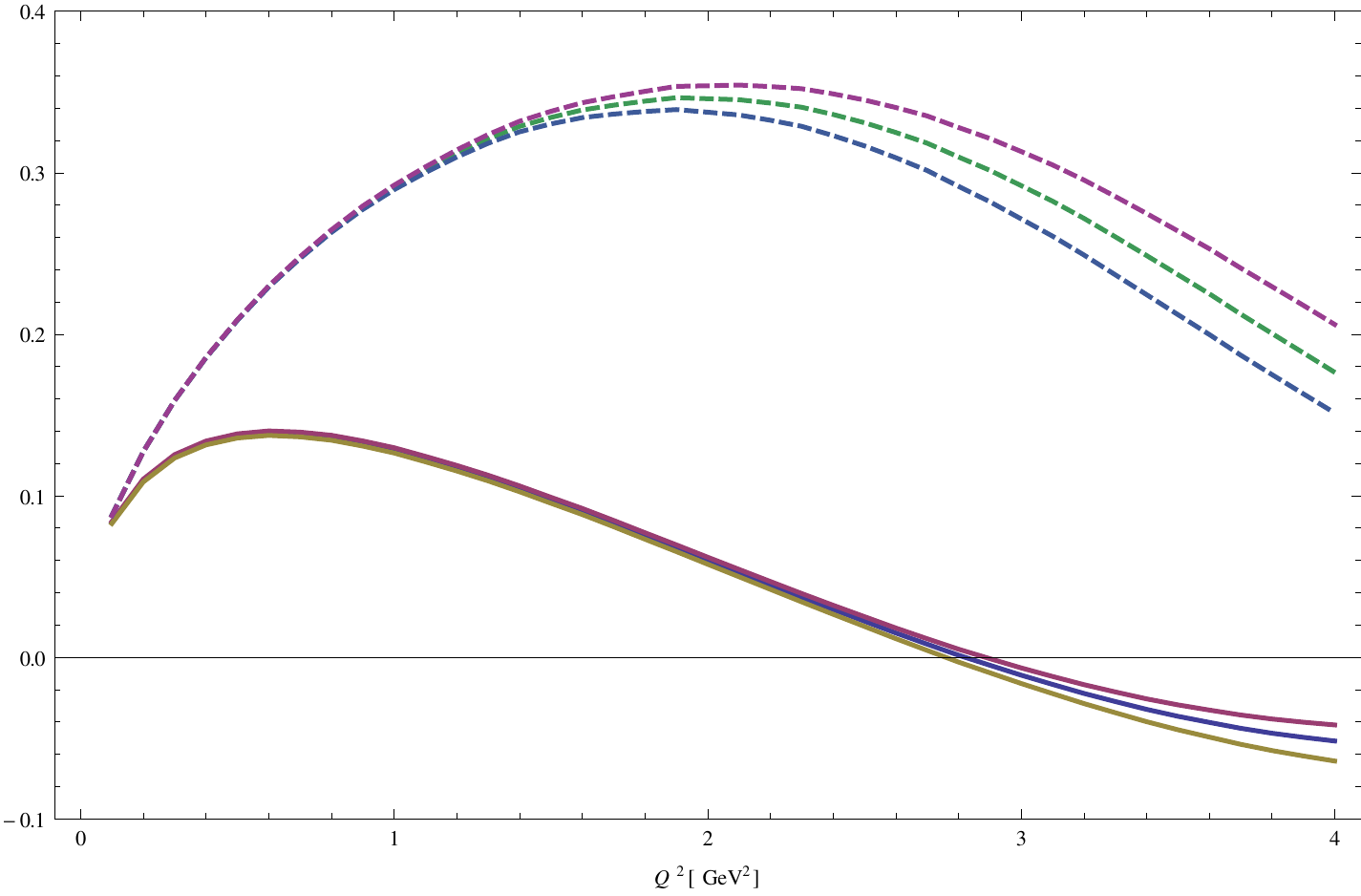} ~~~~~\includegraphics[width=0.45\textwidth]{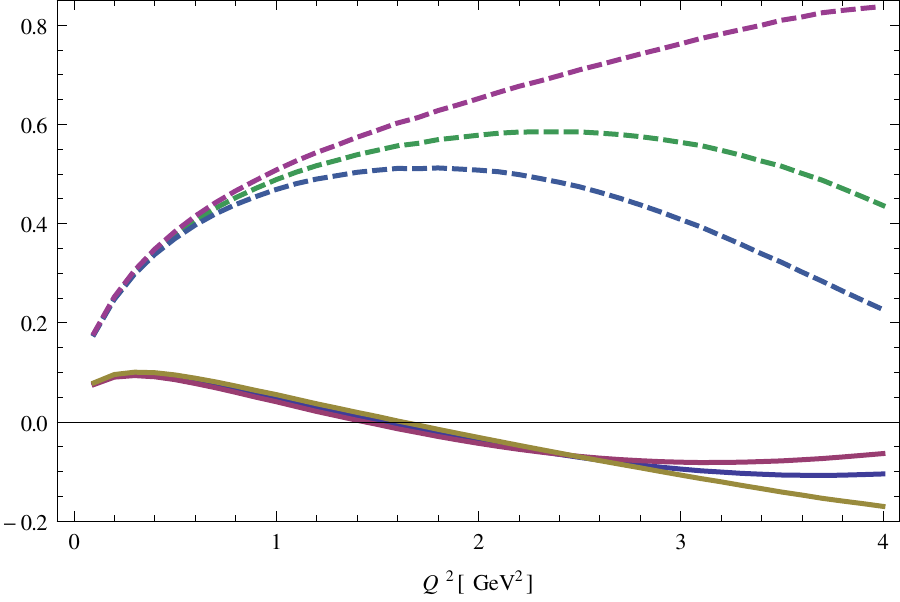}
\caption{The $Q^2$ dependence of the  $<cos~ \varphi> $ (solid curves) and $<sin~ \varphi> $ (dashed curves) moments normalized by the total cross section,  for  $\Delta_T = 0.5$ GeV, $y=0.7$ and $s=20$ GeV$^2$ for a proton (left panel) and a neutron(right panel) target. The  three curves correspond to the three models explained in \cite{Pire:2017lfj}, and quantify the theoretical uncertainty of our estimates.}
   \label{cosPhi-proton}
\end{figure}

$D^+$ neutrino-production allows both quark and gluon GPDs to contribute. Neglecting the strange content of nucleons leads to selecting $d$ quarks in the nucleon, thus accessing the $d$ (resp. $u$) quark GPDs in the proton for the scattering on a proton  (resp. neutron) target, after using isospin relation between the proton and neutron.
The transverse contribution is plotted in Fig. \ref{sigma{--}_dy_D+_on_proton_and_neutron} as a function of $Q^2$  for $y=0.7$ and $\Delta_T=0$. The cross section is reasonably flat in $y$ and $Q^2$ so that an integration over the regions $0.45 < y <1$ and $0.5 < Q^2 < 3$ GeV$^2$ does not require much care.

The longitudinal cross sections dominate the transverse ones, mostly because of the larger values of the chiral-even GPDs, and specifically of the gluonic ones.
The relative importance of quark and gluon contributions to the longitudinal cross sections is shown in Fig. \ref{sigma{00}_dy_D+_on_proton_neutron_quarks_gluons} as a function of $Q^2$ for a specific set of kinematical variables. The $y$ dependence may help to disentangle the longitudinal and transverse contributions since the longitudinal cross section vanishes as $y\to 1$ as is obvious from Eq. (\ref{cs}).

The interference term $\sigma_{-0}$ is accessed through azimuthal moments such as
     \begin{eqnarray}
  <cos ~\varphi>&=&\frac{\int cos ~\varphi ~d\varphi ~d^4\sigma}{\int d\varphi ~d^4\sigma} \approx K_\epsilon\, \frac{{\cal R}e \sigma_{- 0}}{\sigma_{0 0}+K_\epsilon^2 \sigma_{--}}  \,, 
    \label{moments}
   \end{eqnarray} 
   with $K_\epsilon =\frac{\sqrt{1+\varepsilon}+\sqrt{1-\varepsilon} }{2 \sqrt{\epsilon} }$. 
 This allows access to the transversity GPDs in a linear way but requires to consider $\Delta_T \ne 0$ kinematics.
We show on Fig. \ref{cosPhi-proton}  $<cos\varphi>$  and $<sin\varphi>$  for the proton and  neutron target, for the kinematical point defined as $y=0.7, \Delta_T = 0.5$ GeV and $s=20$ GeV$^2$.

In conclusion, let us stress that collinear QCD factorization has allowed us  to calculate exclusive neutrino production of $D-$mesons in terms of GPDs. This reaction gives us a new  experimental access to transversity  GPDs which is quite different from the previous proposals  \cite{transGPDno,transGPDacc}. Planned medium and high energy neutrino facilities \cite{NOVA} and experiments such  as Miner$\nu$a \cite{Aliaga:2013uqz} and MINOS+ \cite{Timmons:2015laq} which have their scientific program oriented toward the understanding of neutrino oscillations  will collect more statistics than presently available and should thus allow  some important progress in the realm of hadronic physics. 

\paragraph*{Acknowledgements.}
\noindent
 This work is partly supported by grant No 2015/17/B/ST2/01838 from the National Science Center in Poland, by the Polish-French collaboration agreements  Polonium and COPIN-IN2P3.


\begin{thebibliography}{99}

  \bibitem{weakGPD}
  B.~Lehmann-Dronke and A.~Schafer,
  Phys.\ Lett.\ B {\bf 521} (2001) 55;
  C.~Coriano and M.~Guzzi,
  Phys.\ Rev.\ D {\bf 71} (2005) 053002;
  P.~Amore, C.~Coriano and M.~Guzzi,
  JHEP {\bf 0502} (2005) 038;
  A.~Psaker, W.~Melnitchouk and A.~V.~Radyushkin,
  Phys.\ Rev.\ D {\bf 75} (2007) 054001.
  
   \bibitem{Kopeliovich:2012dr} 
  B.~Z.~Kopeliovich, I.~Schmidt and M.~Siddikov,
  Phys.\ Rev.\ D {\bf 86}, 113018 (2012) and
 D {\bf 89}, 053001 (2014);
   G.~R.~Goldstein, O.~G.~Hernandez, S.~Liuti and T.~McAskill,
  AIP Conf.\ Proc.\  {\bf 1222} (2010) 248;
B.~Pire, L.~Szymanowski and J.~Wagner,
  arXiv:1705.11088 [hep-ph].

\bibitem{Pire:2017lfj} 
  B.~Pire, L.~Szymanowski and J.~Wagner,
  Phys.\ Rev.\ D {\bf 95}, no. 9, 094001 (2017);
B.~Pire, L.~Szymanowski and J.~Wagner,
  EPJ Web Conf.\  {\bf 112}, 01018 (2016).
  
   \bibitem{PS} 
  B.~Pire and L.~Szymanowski,
  Phys.\ Rev.\ Lett.\  {\bf 115}, 092001 (2015).
  
 

  \bibitem{transGPDno}
  M.~Diehl {\em et. al.}
  Phys.\ Rev.\  D {\bf 59}, 034023 (1999);
  J.~C.~Collins {\em et. al.},
  Phys.\ Rev.\  D {\bf 61}, 114015 (2000).

\bibitem{transGPDacc}
 D.~Yu.~Ivanov  {\it et al.},
  Phys.\ Lett.\  B {\bf 550}, 65 (2002);
R.~Enberg {\em et. al.},  
  Eur.\ Phys.\ J.\  C {\bf 47}, 87 (2006);
 M.~El~Beiyad  {\it et al.},
 Phys.\ Lett.\ B {\bf 688}, 154 (2010);
   S.~V.~Goloskokov and P.~Kroll,
  Eur.\ Phys.\ J.\ A {\bf 47}, 112 (2011);
  R.~Boussarie {\em et. al.},
  JHEP {\bf 1702}, 054 (2017).


 \bibitem{NOVA} 
  D.~S.~Ayres {\it et al.}  [NOvA Collaboration],
  hep-ex/0503053; see also
   M.~L.~Mangano, S.~I.~Alekhin, M.~Anselmino {\it et al.},
  CERN Yellow Report CERN-2004-002, pp.185-257
  [hep-ph/0105155].


\bibitem{Aliaga:2013uqz} 
  L.~Aliaga {\it et al.} [MINERvA Collaboration],
  Nucl.\ Instrum.\ Meth.\ A {\bf 743}, 130 (2014);
    C.~L.~McGivern {\it et al.} [MINERvA Collaboration],
  Phys.\ Rev.\ D {\bf 94}, no. 5, 052005 (2016).
  
\bibitem{Timmons:2015laq} 
  A.~Timmons,
  Adv.\ High Energy Phys.\  {\bf 2016}, 7064960 (2016).
\end{thebibliography}
\end{document}